\documentclass[aps,jpc,showpacs,twocolumn,superscriptaddress]{revtex4-2}
\usepackage{epstopdf}
\usepackage{graphicx}
\usepackage{dcolumn}
\usepackage{bm}

\usepackage[pdfpagemode=UseNone,pdfstartview=FitH,colorlinks=true,linkcolor=blue,urlcolor=blue,anchorcolor=blue,citecolor=blue]{hyperref}
\usepackage{cancel,xcolor}
\usepackage{soul,xcolor}
\setstcolor{red}

\usepackage[utf8]{inputenc}
\usepackage[T1]{fontenc}
\usepackage{mathptmx}
\usepackage{etoolbox}

\usepackage{comment}
\usepackage{braket}
\usepackage{physics}

\usepackage{color}

\def\sig{{\mbox{\boldmath{$\sigma$}}}}
\def\nab{{\mbox{\boldmath{$\nabla$}}}}
\def\om{{\mbox{\boldmath{$\omega$}}}}

\begin{document}

\preprint{AIP/123-QED}

\title{Spin-orbit interactions,  time-reversal symmetry, and spin selection}


\author{Amnon Aharony}
\email{aaharonyaa@gmail.com}
\affiliation{School of Physics and Astronomy, Tel Aviv University, Tel Aviv 6997801, Israel}

\author{Ora Entin-Wohlman}
\email{orawohlman@gmail.com}
\affiliation{School of Physics and Astronomy, Tel Aviv University, Tel Aviv 6997801, Israel}
\date{\today}


\begin{abstract}
Spin selective transport is usually associated with spin-orbit interactions. 
However, these interactions are invariant under time-reversal symmetry, and the Onsager relations and Bardarson's theorem imply that such interactions cannot yield spin selectivity for transport through a junction between two electronic reservoirs. Here, we review several ways to overcome this restriction, using a Zeeman magnetic field, the Aharonov-Bohm phase, time-dependent electric fields that generate time-dependent spin-orbit interactions, 
time-dependent transients, more than two terminals, leakage,
and more than one level per ion on the junction.  Our considerations focus on the transport of noninteracting electrons at low temperatures. A possible connection with the phenomenon of chiral-induced spin selectivity (CISS) is pointed out in one of the systems considered.

\end{abstract}


\maketitle

\section{Introduction}
{\label{intro}


Quantum computers use qubits, which are two-component spinors, of the form $|\chi\rangle=\big(\cos(\theta/2),\sin(\theta/2)e^{i\phi}\big)^T$ with the elevation and azimuthal angles $\theta$ and $\phi$. Quantum information (i.e., the real numbers $\theta$ and $\phi$) can be written on these qubits and then be manipulated (or read) in quantum networks. Here, we discuss mobile qubits, representing the (spin $1/2$)  states of electrons. Each such spinor is an eigenstate of the electron's magnetization along the unit vector $\hat{\bf m}=(\sin\theta\cos\phi,\sin\theta\sin\phi,\cos\theta)$, i.e. an eigenstate of the $2\times 2$ matrix $\sig\cdot\hat{\bf m}$, where $\sig$ is the three-component vector of the Pauli matrices.  As electrons move through a spintronic device, their magnetizations may rotate, thus changing the quantum information. In the ideal case of spin filters, arbitrary incoming spins all exit with the same (tunable) magnetization, therefore allowing quantum information to be written on them. Combining such filters also allows reading that information~\cite{filter}.


A natural way to manipulate electronic spins uses the spin-orbit interaction (SOI), which has the form
\begin{align}
{\cal H}^{}_{\rm so}\propto {\bf E}\cdot[{\bf p}\times\sig]\ ,
\end{align}
where ${\bf E}=-\nab V\equiv E\hat{\bf n}$ is the electric field ($V$ is the electric potential) and ${\bf p}\equiv \hbar {\bf k}$ is the electron's momentum. Below, we set $\hbar=1$. For linear or planar networks, we also choose  $\hat{\bf n}=\hat{\bf z}$. 
 As Rashba showed~\cite{rashba}, the electric potential $V$, which restricts the motion of an electron to the two-dimensional $XY$ plane, generates a spin-dependent potential,
\begin{align}
{\cal H}^{}_{\rm so}=\alpha[{\bf p}\times\sig]^{}_z\ ,
\end{align}
where $\alpha =k^2_{\rm so}/(2m^\ast)\propto E$ measures the SOI strength ($k^{}_{\rm so}$ measures this strength in momentum units, and $m^\ast$ is the effective mass of the electronic band). The total Hamiltonian of the (free) electron can then be written as
\begin{align}
{\cal H}=\frac{1}{2m^\ast}\big[{\bf p}+k^{}_{\rm so}\sig\times\hat{\bf n}\big]^2\ .
\end{align}
This reflects an effective shift of the momentum, which implies a rotation of the `free' electron spinor, $e^{i{\bf k}\cdot {\bf R}}|\chi\rangle$ into ~\cite{entin} 
\begin{align}
e^{i{\bf k}\cdot {\bf R}}|\chi\rangle \Rightarrow e^{i(k^{}_{\rm so}[\hat{\bf n}\times\sig]+{\bf k})\cdot {\bf R}}|\chi\rangle\ .
\label{entin}\end{align}
This effect, sometimes also called the `Aharonov-Casher' effect~\cite{AC}, led to the proposal of the Datta-Das spin field-effect transistor~\cite{DD}.
The Aharonov-Casher phase is similar to the Aharonov-Bohm and Aharonov-Ananden topological phases~\cite{ABC}.


In this paper, we consider mesoscopic spintronic devices that are placed between two electronic reservoirs. A major constraint on spin-resolved transport between
two terminals arises from Bardarson’s theorem~\cite{bardarson}: spin selectivity is forbidden for a two-terminal time-reversal symmetric system. Choosing a basis in which the reflected wave in each reservoir is the time-reversed incoming wave, Bardarson proved that, with time-reversal symmetry, the scattering matrix $S$ must obey $S^T=-S$, and therefore, the transmission and reflection matrices are proportional to the unit matrix, and the scattering does not change the spin.  Since the SOI is time-reversal-invariant, it cannot by itself generate spin splitting in such simple devices~\cite{com}.
Below, we list various ways 
 to overcome this restriction and obtain spin selectivity (or filtering). These include a Zeeman magnetic field, time-dependent electric fields that generate time-dependent spin-orbit interactions, 
time-dependent transients, more than two terminals, leakage,
and more than one level per ion on the junction. All of these methods exploit quantum interference, which has a destructive effect on one spin direction and is constructive for the other.


Our paper concentrates only on the simplest situation of spin selectivity in electron transport at very low temperatures and systems in which electron-electron interactions are assumed to be small. 
For the special case
  of {\it chiral-induced} spin selectivity (CISS) there exist papers that add electron-electron and electron-phonon interactions to explain the breakdown of time-reversal symmetry~\cite{int,CISSrev}. Some of these models have also been supported by experiments~\cite{eph}. The optical activity due to the coupling to external electromagnetic fields has also been found to generate CISS~\cite{optic}. These extensions go beyond the scope of the present review. Below we concentrate on simple mesoscopic structures.  In one case, we mention a possible connection to the CISS effect.

 Section \ref{II} describes the tight-binding model, which we impose on discrete quantum networks. The following sections then apply the various methods to 
 a weak link (\ref{weak}), an Aharonov-Bohm-Rashba interferometer (\ref{ABR}), and a helical molecule (\ref{CISS}).
 

\section{Tight-binding model}
\label{II}


Our calculations have been done on tight-binding network models~\cite{comtrans}. For an eigenstate with energy ${\cal E}$, the tight-binding equation for the (two-component) spinor $|\psi(i)\rangle$ at site $i$ of the network is
\begin{align}
({\cal E}-\epsilon^{}_i)|\psi(i)\rangle=-\sum^{}_j J^{}_{ij}U^{}_{ij}|\psi(j)\rangle\ ,
\label{tb}
\end{align}
where $J^{}_{ij}$ is the `bare' transmission energy of the hopping between the two sites, and $U^{}_{ij}$ is a $2\times 2$ unitary matrix, which rotates the spinor on which it acts. For a pure SOI,  the tunneling amplitude through each bond has the form of Eq. ~(\ref{entin}). With a magnetic flux, this becomes
\begin{align}
J^{}_{ij} U^{}_{ij}=J^{}_{ij}\exp[ik^{}_{\rm so}[\hat{\bf d}\times \hat{\bf n}]\cdot\sig L-i\phi^{}_{ij}]\ ,
\label{amp}
\end{align}
where $\hat{\bf d}$ is a unit vector along the bond between sites $i$ and $j$, $L$ is the bond length, the electric field ${\bf E}$ is along the unit vector 
$\hat{\bf n}$, and $\phi^{}_{ij}$ is the (partial) Aharonov-Bohm (AB) phase associated with the magnetic field perpendicular to the $i-j$ bond. The results depend only on the sums of these phases around loops in the network and on the Aharonov-Casher phases.
For a network of $N$ sites, we have to solve the $2N$ equations~(\ref{tb}).


\section{Weak links}
\label{weak}


The simplest device connecting two electronic reservoirs is a single (one-dimensional) weak link, described by a tunneling matrix between the reservoirs. In the general case, this matrix is proportional to the spin-dependent propagator connecting its two ends~\cite{Shahbazyan1994}. Simpler cases are described in the previous section. Since the SOI alone on the single weak link between two reservoirs is insufficient for spin splitting, we must complement it with more ingredients.


\subsection{Zeeman field}

A simple, direct way to break time-reversal symmetry is to apply a Zeeman magnetic field on the (one-dimensional) weak link~\cite{Sh1}. 
In conjunction with the Rashba SOI, such a field does yield a non-zero spin current, whose magnitude and direction depend on the direction of the field. When this field is not parallel to the effective field due to the SOI, which is along $\hat{\bf n}\times\hat{\bf k}$ (perpendicular to the electric field and the direction of motion), both the charge and the spin currents oscillate with the length of the wire. Measuring the oscillating anisotropic magnetoresistance can thus yield information on the SOI strength. These features can be tuned by varying the magnetic and/or electric field, with possible applications to spintronics.


\subsection{Zeeman fields and reservoir magnetizations}

A Datta-Das spin field-effect transistor~\cite{DD} is built of a heterostructure with a Rashba spin-orbit interaction at a weak link  (or quantum well) separating two possibly magnetized reservoirs. The particle and spin currents between the two reservoirs are driven by chemical potentials that are (possibly) different for each spin direction. These currents can be tuned by varying the strength of the SOI, which changes the amount of the spin rotation of electrons crossing the link. The currents depend on additional Zeeman fields on the link and on variations of the reservoir magnetizations~\cite{sarkar}. In contrast to the particle current, the spin currents are not necessarily conserved; an additional spin polarization is injected into the reservoirs. If a reservoir has a finite (equilibrium) magnetization, then the spin current into that reservoir can only have spins that are parallel to the reservoir magnetization, independent of all the other fields. This spin current can be enhanced by increasing the magnetization of the other reservoir and can also be tuned by the SOI and the various magnetic fields. When only one reservoir is magnetized, then the spin current into the other reservoir has an arbitrary tunable size and direction. In particular, this spin current changes as the magnetization of the other reservoir is rotated. The optimal conditions for accumulating spin polarization on an unpolarized reservoir are to either apply a Zeeman field in addition to the SOI or to polarize the other reservoir.


\subsection{Magnetoconductance}

The Aharonov-Casher phase depends on both the device geometry and the SOI strength. This phase can be directly found by measuring the anisortropy of the magnetoconductance of an SOI active weak link~\cite{MC}.  Measuring this magnetoconductance anisotropy thus allows the calibration of Rashba spintronic devices by an external electric field that tunes the spin-orbit interaction and, hence, the Aharonov-Casher phase.


\subsection{Time-dependent SOI}

Another simple way to break time-reversal symmetry is to introduce an explicit time-dependence of the Hamiltonian. For example, 
 an AC electric field, that rotates periodically in the plane perpendicular to the link, generate a time-dependent SOI that injects spin-polarized electrons into the terminals~\cite{DC}. The injected spin polarization has a DC component (independent of time) along the link and a rotating (time-dependent) transverse component in the perpendicular plane. In the low-rotation-frequency regime, these polarization components are proportional to the frequency. The DC component of the polarization vanishes for a linearly polarized electric field.


\section{The  Aharonov-Bohm-Rashba interferometer}
\label{ABR}


Many measurements of quantum phases are based on the interference between two paths in interferometers. 
For the planar diamond interferometer in Fig.~\ref{AB1}, with both an AB flux and an SOI, one can eliminate the sites $b$ and $c$ from the tight-binding equations ~(\ref{tb}) and end up with an effective non-unitary tunneling matrix amplitude between the site $0$ and $1$ \cite{filter}, 
\begin{align}
W&=\gamma^{}_bU^{}_{0b}U^{}_{b1}+\gamma^{}_cU_{0c}U^{}_{c1}\ ,\nonumber\\
 \gamma^{}_{b}&=J^{}_{0b}J^{}_{b1}/({\cal E}-\epsilon^{}_{b})\ , \ \ \gamma^{}_{c}=J^{}_{0c}J^{}_{c1}/({\cal E}-\epsilon^{}_{c})\ ,
\label{W}
\end{align}
where $\epsilon^{}_{b,c}$ is the site energy on these sites. The unitary matrices $U^{}_{ij}$ are given by Eq.~(\ref{amp}).


\subsection{Spin filters}


Since the tunneling amplitude $W$ in Eq. (\ref{W}) is a non-unitary matrix, the interference between the two paths can become destructive~\cite{filter}. Specifically, the transmission matrix from the left lead to the right one depends on $W$, and its two eigenvalues thus depend on the (total) Aharonov-Bohm phase $\Phi$ and the Aharonov-Casher phase $\omega$, associated with the spin-dependence of the product $U^{}_bU^{\dagger}_c\propto \exp[-i\Phi+i\om\cdot\sig]$, representing the hopping of the electron around the loop. An appropriate choice of the electric and magnetic fields, i.e., the Aharonov-Casher and Aharonov-Bohm phases, can generate a full destructive interference of one spin state and, thus, a full polarization of the other spin state, i.e., full spin filtering. This tunable, completely polarized spin state exits the interferometer for any arbitrary incoming spin state \cite{filter}. Combining two such interferometers also allows reading the quantum information on the flying qubits.


\begin{figure}
\vspace{-1cm}
\includegraphics[width={8.5cm}]{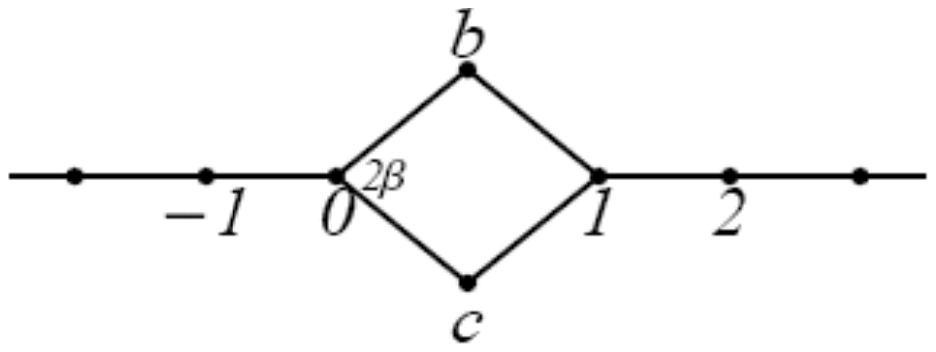}
\vspace{-8.5cm}
\caption{
 Diamond interferometer. }
\label{AB1}
\end{figure}


 Similar results also apply to an interferometer made of two elongated quantum dots or quantum nanowires, which are subjected to both an Aharonov–Bohm flux and (Rashba and Dresselhaus \cite{dresselhaus1955}) spin-orbit interactions~\cite{tarucha}. Such a double-dot interferometer can serve as a perfect spin filter due to a spin interference effect. By appropriately tuning the external electric and magnetic fields, which determine the Aharonov–Casher and Aharonov–Bohm phases, and with some relations between the various hopping amplitudes and site energies, the interferometer blocks electrons with a specific spin polarization, independent of their energy. The blocked polarization and the polarization of the outgoing electrons are controlled solely by the external electric and magnetic fields and do not depend on the energy of the electrons. Furthermore, the spin filtering conditions become simpler in the linear-response regime, in which the electrons have a fixed energy. Unlike the diamond interferometer, spin filtering in the double-dot interferometer does not require high symmetry between the hopping amplitudes and site energies of the two branches of the interferometer and, thus, may be more appealing from an experimental 
point of view.


 \subsection{Real-time Transients effects}
 \label{trans}
 
 Starting with an arbitrary initial occupation of the sites on the interferometer, the quantum states of the system have a transient time dependence until it reaches the time-independent stationary state~\cite{weimin}. In the transient-time regime, the system does not possess time-reversal symmetry and thus can support a spin-dependent current through the interferometer, even for 
initially unpolarized spins. Then,  there appear spin currents in the electrodes, for which the spins in each electrode are polarized along characteristic directions, predetermined by the SOI parameters and by the geometry of the system. Special choices of the system parameters yield steady-state currents in which the spins are fully polarized along these characteristic directions. The time required to reach this steady state depends on the couplings of the double quantum dots to the leads. The magnitudes of the currents depend strongly on the SOI-induced effective fluxes. Without the magnetic flux, the spin-polarized current cannot be sustained to the steady states due to the phase rigidity for this system. For a nondegenerate double quantum dot, transient spin transport can be produced by the sole effects of SOI. Interestingly, one finds that the spin-resolved currents can be extracted from measurements of the total charge current.


\subsection{Several terminals}\label{more}


As stated, the Bardarson theorem \cite{bardarson} forbids spin polarization in two-terminal devices. 
In contrast, a spin polarization can be generated when electrons from one source reservoir flow into two (or more) separate drain reservoirs~\cite{3t}. When the electrons are transmitted through a “diamond” interferometer into two drains, they can be simultaneously fully spin-polarized, along different tunable directions, even when the two arms of the interferometer are not identical.


\subsection{Several orbital states}
\label{utsumi}


For a single-orbital state on each site of the network, the Bardarson theorem \cite{bardarson} is based on the degeneracy of the two spin states in the transmission (or the reflection) $2\times 2$ matrices for each tunneling channel. However, the theorem does not apply when the system consists of atoms having more than a single orbital. 
Indeed, when each atom has two orbital states, then the transmission and reflection matrices become of order $4\times 4$ for each channel, and spin-resolved currents can be generated without breaking time-reversal symmetry~\cite{utsumi}. 
In this case, spin-flip processes can be accompanied by a flip of the orbital channels.  We come back to this type of models for spin selectivity in the next section.


\section{Helical  molecules and the CISS effect}
\label{CISS}


A special case of spin splitting, which has drawn much interest during the last two decades, involves the chiral-induced spin selectivity (CISS) effect~\cite{naaman}.  The CISS effect refers to a spin transport phenomenon that is specific to chiral molecules:
When an electron is injected into chiral molecules such as DNA, the electron spin is selectively separated, depending on the chirality.
This phenomenon exhibits a high spin polarization rate compared to ordinary magnetic materials and has attracted attention as a quantum phenomenon that occurs at room temperature. 
Although the CISS effect has been observed in many experiments using various chiral materials, a fully convincing theoretical explanation has not yet been proposed~\cite{CISSrev}.
Attempts to explain this phenomenon using only the spin-orbit interactions were not successful due to Bardarson's theorem~\cite{bardarson,com}. Below, we describe some of the methods mentioned in Secs. \ref{weak} and \ref{ABR} to overcome the outcome of Bardarson's theorem to the CISS case. To this end, the molecule is described by sites on a helical chain, coupled by the tunneling amplitude $J$; interference is introduced by adding tunneling $\tilde{J}$ between sites that "sit" on top of each other on the helix, see Fig. \ref{helix} and Ref. \cite{utsumi}. The model exhibits spin
splitting without breaking time-reversal symmetry: the intra-atomic SOI induces concomitant spin and orbital
flips. 


A companion model to explore the effect of chiral molecules exploits the Edelstein effect \cite{Edelstein1990} that converts charge to angular momentum. 
The orbital angular momentum is directly generated by the chirality of the crystal. The spin-splitting band relies on the SOI.
A recent publication \cite{Gobel2025} combines these two features to describe the
chirality-induced orbital selectivity (termed CIOS), showing that the orbital contribution surpasses the spin contribution by orders of magnitude.


\begin{figure}
\vspace{-1.5cm}
\includegraphics[width={9cm}]{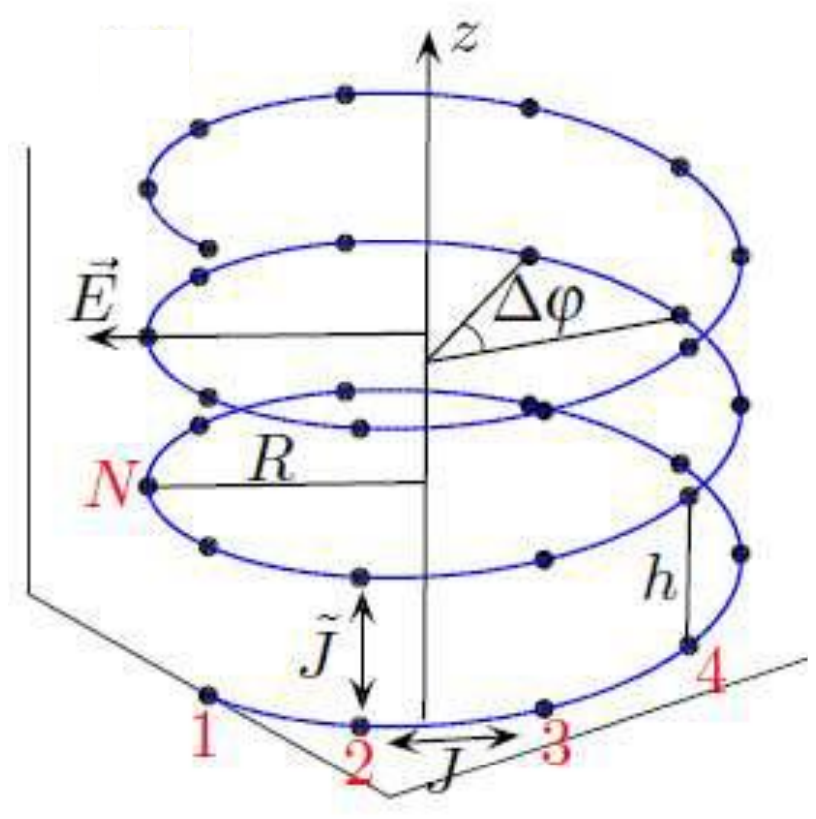}
\vspace{-6.cm}
\caption{
Schematic picture of a single strand of a double-
stranded DNA. $R(\phi^{}_{n})$  is the radius vector of the $n$th site within the Frenet-
Serret scheme, $\Delta\phi=2\pi/N$,  $ h$ is the pitch, and $\phi^{}_{n}=n\Delta\phi$.}
\label{helix}
\end{figure}


\subsection{Transients} 

In Ref.~\cite{CISStrans}, the application of a pulse gate voltage between the ends of chiral molecules generated spin-polarized electrons on the substrate under these molecules. We conjecture that this transient magnetization results from the non-equilibrium time-dependent spin transport through the molecules, similar to that calculated in Sec. \ref{trans}.


\subsection{Several terminals and leakage}


Using the same techniques as in Sec. \ref{more},  a single helical molecule attached to more than one drain can induce a significant spin polarization in electrons passing through it~\cite{3t}. The average polarization remains nonzero even when the electrons outgoing into separate leads are eventually mixed incoherently into one absorbing reservoir. This may explain recent experiments on the spin selectivity of certain helical-chiral molecules~\cite{sphere}. In these experiments, the helical molecule was connected to a metallic sphere, and polarized electrons were detected by an AFM tip on the sphere. This polarization can be because the sphere detected electrons coming out from more than one site at the end of the molecule.

Spin selection can also be generated if electrons can flow out of each site on the helix, into the outside environment, indicating a breakdown of unitarity, or `leakage'~\cite{mat,balseiro}. Several of the experiments use a two-dimensional array of parallel molecules, and we conjecture that electrons may leak out of each molecule into the neighboring molecules.


 \subsection{Several orbital states}
 
 The techniques described in Sec. \ref{utsumi} have also been applied to a helical molecule~\cite{utsumi}, using a two-orbital tight-binding model similar to Fig. \ref{helix} with intra-atomic spin-orbit interactions.  One way to obtain two orbital states on each site of the helix is to include the three $p$ orbitals of carbon atoms and add a strong crystalline field along the helix~\cite{utsumi1}.  The typical energy scale of the helical states is the product of the intra-atomic SOI and the curvature. The spin filtering mechanism is associated with the intra-atomic SOI, which would be larger than the inter-atomic SOI. In this respect, this model may be a more likely candidate for the CISS in organic material than other models associated with the inter-atomic SOI. Spin-up and -down states propagate in opposite directions, leading to the CISS effect. 
Such pairs of orbital states may also arise in double helices, like those used experimentally in Ref.~\cite{sphere}.


 The above model also produces the enhancement of charge modulations, concentrated at the edges of the molecule,  due to the evanescent states~\cite{utsumi2}. A Zeeman field at an edge of the atomic chain, which breaks the time-reversal symmetry, yields a finite spin polarization whose direction depends on the chirality of the molecule. The chirality change induces a reasonable amount of the energy difference, which may provide an insight into the enantio-selective adsorption of chiral molecules on the ferromagnetic surface.


\section{Synopsis}


The chiral structure of molecules like DNA and of crystals 
like crystalline tellurium and selenium affects electronic transport: The chirality of a structure determines the sign of the spin-polarization of electrons moving through it. This innate ability of chiral materials to act as spin filters reveal the fundamental connection
between chiral symmetry and electron spin and promise profound and ubiquitous implications for existing technologies like spintronics, and new approaches, for instance,  to answer the intriguing appearance of the homochiral nature of life.


However, spin polarization brought about by the tunneling of electrons through spin-orbit active weak links and tunnel junctions carries out wider promises. In particular, tuning spin-dependent phenomena without applying a magnetic field, for instance, by exploiting the Rashba spin-orbit coupling created by an electric field, makes
it possible for the electronic spin to couple to an electric field. Put another way, the magnitude of an electronic property arising from the interfacial
breaking of inversion symmetry can be
modulated by applying an external electric field \cite{Caviglia2010}.


Spin polarization of the electrons, occurring as electrons tunnel through an SOI-active weak link, is expected to allow for an electrically generated magnetization of a small-size non-magnetic device. 
However, the constraint originating from the time-reversal symmetry of the SOI prohibits any effect of it on the two-terminal tunneling of electrons in non-superconducting devices \cite{bardarson}.  In our short review, we have highlighted several (theoretical) approaches to alleviate this problem and pointed out their connections to existing experiments.
We hope that the above list of possibilities will stimulate more experiments and further analyses of existing experiments.


The work reviewed here was done with many collaborators, including Shlomi Mattiyahu,  Yasuhiro Tokura, 
Seigo Tarucha, Shingo Katsumoto,
Robert Shekhter, Mats Jonson,
Wei-Min Zhang and  Carlos Balseiro. Especially, we thank Yasuhiro Utsumi for his very fruitful recent collaborations.


Over the years, we have had many discussions with Abraham Nitzan. We are happy to participate in celebrating his many contributions to chemical physics.




\begin{thebibliography}{99}
\bibitem{comtrans} As mentioned in Ref. \onlinecite{int}, the electronic molecular orbitals in some realistic organic molecules may break the translational symmetry assumed in our tight-binding model.
\end{thebibliography}
\end{document}